\documentclass[10pt]{article}
\usepackage[OE]{express}

\begin{document}
\title{An extended model of the quantum free-electron laser}

\author{M.~S.~Brown,\authormark{1,5} J.~R.~Henderson,\authormark{1,2,3,4,6} 
L.~T.~Campbell,\authormark{1,2,3,7} and B.~W.~J.~McNeil\authormark{1,2,3,*}}

\address{
\authormark{1}SUPA, Department of Physics, University of Strathclyde, Glasgow, G4 0NG, UK\\
\authormark{2}ASTeC, STFC Daresbury Laboratory, Warrington, WA4 4AD, UK\\
\authormark{3}Cockcroft Institute, Warrington, WA4 4AD, UK\\
\authormark{4}Department of Engineering, Lancaster University, LA1 4YR, UK}

\email{\authormark{5}m.s.brown@strath.ac.uk}
\email{\authormark{6}james.henderson@cockcroft.ac.uk}
\email{\authormark{7}lawrence.campbell@strath.ac.uk}
\email{\authormark{*}b.w.j.mcneil@strath.ac.uk}



\begin{abstract}
Previous models of the quantum regime of operation of the Free Electron Laser (QFEL) have performed an averaging and the application of periodic boundary conditions to the coupled Maxwell -- Schr\"odinger equations over short, resonant wavelength intervals of the interaction. Here, an extended, one-dimensional model of the QFEL interaction is presented in the absence of any such averaging or application of periodic boundary conditions, the absence of the latter allowing electron diffusion processes to be modeled throughout the pulse. The model is used to investigate how both the steady-state (CW) and pulsed regimes of QFEL operation are affected. In the steady-state regime it is found that the electrons are confined to evolve as a 2-level system, similar to the previous QFEL models. 
In the pulsed regime Coherent Spontaneous Emission (CSE) due to the shape of the electron pulse current distribution is shown to be present in the QFEL regime for the first time. However, unlike the classical case, CSE in the QFEL is damped by the effects of quantum diffusion of the electron wavefunction. Electron recoil from the QFEL interaction can also cause a diffusive drift between the recoiled and non-recoiled parts of the electron pulse wavefunction, effectively removing the recoiled part from the primary electron-radiation interaction.
\end{abstract}



\section{Introduction}
The Free-Electron Laser (FEL) is a unique and powerful scientific instrument capable of generating coherent radiation across a wide range of the electromagnetic spectrum into the hard X-ray~\cite{lcls,sacla,euxfel,np}. The FEL uses a  beam of relativistic electrons  which is propagated through an undulator, usually consisting of a sinusoidally varying static magnetic field or possibly a counter-propagating high intensity coherent laser field. The relativistic electrons, with a mean Lorentz factor of $\gamma_0$, are forced by the undulator to oscillate transverse to their direction of propagation and emit electromagnetic radiation at a resonant wavelength $\lambda_r=\lambda_u(1+\bar{a}_u^2)/2\gamma_0^2$, where $\lambda_u$ is the undulator period and $\bar{a}_u\propto B_u,\lambda_u$, with $B_u$ the undulator magnetic field strength. The undulator parameter determines the electron oscillation amplitude~\cite{np}. In the case of a laser-undulator~\cite{lasund}, typically of much shorter period $\lambda_u$ than that of a permanent magnet undulator, the resonance relation is $\lambda_r=\lambda_u(1+\bar{a}_u^2)/4\gamma_0^2$, where now the undulator parameter $\bar{a}_u$ is proportional to the laser undulator intensity~\cite{oc17}.

A collective instability between the oscillating electrons and the radiation they emit can cause the electrons to bunch along their direction of propagation at the resonant wavelength and to emit coherently. The increased radiation field further bunches the electrons and closes a positive feedback loop, resulting in a high gain, exponential amplification of the radiation field and bunching of the electrons until non-linear over-bunching of the electrons saturates the emission process. 

The original description of the FEL interaction used a quantum mechanical description in the small gain regime~\cite{Madey}. It was later shown that the exponentially unstable high gain FEL regime could be described using a purely classical  framework~\cite{hopf,colson,np} starting from the intrinsic shot-noise  within the electron beam in a process called Self-Amplified Spontaneous Emission (SASE). This has seen the realization of FELs generating coherent radiation into the hard X-ray without the need for seed lasers. However, for sufficiently short wavelength scales, the single photon emission recoil of an electron can be sufficiently large that the discrete nature of photon emission cannot be neglected~\cite{Schiavi,Bonifacio_harmonics,friedman,P_Kling,P_Preiss} and a quantum description, at least of the electron dynamics of the FEL interaction, becomes necessary.  Hence, a return has been made to a quantum description of the FEL, called a QFEL~\cite{Bonifacio_coherent,Piovella_quantum_FEL,Schroeder}, which can better describe the generation of very short wavelength radiation potentially into the gamma ray region of the spectrum. Clearly, such coherent output would prove a unique tool, e.g.~in the study of nuclear processes, and has the potential to open up many new frontiers across science.

All previous QFEL models have averaged the coupled Maxwell -- Schr\"odinger equations describing the radiation/electron interaction over short intervals of one resonant wavelength within the system. Periodic boundary conditions were also applied to the electron wavefunction over the same interval which inhibits electron diffusion throughout the pulse. These approximations are also commonly applied in classical models which use the Maxwell -- Lorentz equations. However, in such classical models, effects such as Coherent Spontaneous Emission~\cite{doria,CSE1} cannot be modeled, and radiation emission is restricted to a narrow bandwidth around the defined resonant frequency. The application of periodic boundary conditions to the electron motion means the beam current profile on the order of the resonant wavelength is assumed static, so electron drift between the different regions of the beam is usually neglected. In the quantum regime of operation, the electrons must - by definition - recoil to energy levels outside the usual classical FEL bandwidth, and so the electron drift effects due to this recoil may play a more important role in the QFEL interaction, so requiring a revised and extended QFEL model without averaging or the application of localized periodic boundary conditions to the electron wavefunction.
    
Given these considerations, it would seem prudent to investigate the QFEL interaction using such an extended  QFEL model, and this is the primary topic of this paper. Relatively simple systems, e.g. the absence of electron shot-noise or energy spread, are modeled to allow for a comparison with, and demonstration of, the differences between the previous and extended QFEL models.
Furthermore, we do not limit the parameters used in the simulations to those currently achievable in the laboratory, but rather explore, for the first time, the underlying physics of the extended QFEL model.

\section{The extended 1-D QFEL model}
To  model the semi-classical regime of the FEL~\cite{Preparata,Bonifacio_propagation}, the classical description of a distribution of discrete electrons is replaced by a quantum-mechanical description  of the electrons via a single wavefunction, $\psi$. This electron wavefunction is coupled self-consistently to a scaled classical radiation field, $A$. The scaling used is the classical scaling of~\cite{BNP} where the classical FEL parameter is 
$\rho_F=\gamma_r^{-1} (\bar{a}_u\omega_p/4ck_u)^{2/3} $.
Using this semi-classical model in the absence of any localized periodic boundary conditions, the unaveraged QFEL system can be written in terms of the scaled, coupled Schr\"odinger-Maxwell equations in the 1D limit as~\cite{Bonifacio_propagation}: 
\begin{equation}
\label{schrod1}
i\frac{\partial \psi(\bar{z}, \bar{z}_1)}{\partial \bar{z}} = 
-4\frac{\rho_{F}^2}{\overline{\rho}}\frac{\partial ^2 \psi(\bar{z}, \bar{z}_1)}{\partial \bar{z}_1^2} 
-i\frac{\overline{\rho}}{2}\bigg[A(\bar{z}, \bar{z}_1)e^{i\frac{\bar{z}_1}{2\rho_F}} - c.c \bigg]\psi(\bar{z}, \bar{z}_1), 
\end{equation}

\begin{equation}
\label{wave1}
\bigg[\frac{\partial}{\partial \bar{z}} + \frac{\partial}{\partial \bar{z}_1} \bigg]A(\bar{z}, \bar{z}_1)
=|\psi(\bar{z}, \bar{z}_1)|^2e^{-i\frac{\bar{z}_1}{2\rho_F}} + i\frac{\delta}{\overline{\rho}}A(\bar{z}, \bar{z}_1) ,
\end{equation}
where $\bar{z} = z/l_g$ is the scaled undulator distance in units of the gain length $l_g=\lambda_u / 4\pi \rho_F$, and $\bar{z}_1=(z-c \bar{\beta}_{z}  t)/\bar{\beta}_{z}l_c$, is a localised coordinate along the electron bunch in units of the cooperation length, $l_c=\lambda_r / 4\pi \rho_F$. The electron energy detuning from resonance is $\delta=2mc (\gamma_0-\gamma_r) / (\bar{h} k ) $. 
The quantum FEL parameter is defined as:
\begin{equation}
\bar{\rho} = 2\bigg(\frac{mc\gamma_r}{\hbar k}\bigg)\rho_F ,
\end{equation}
and is proportional to the ratio between the saturated momentum spread of the electron beam due to the classical FEL interaction, $\gamma_{r}mc\rho_F$, and the single photon recoil momentum, $\hbar k$. The value of the quantum FEL parameter $\bar{\rho}$ governs the transition between the classical $(\bar{\rho}\gg 1)$ and quantum $(\bar{\rho}\lesssim 1)$ regimes of operation~\cite{Bonifacio_noise}. Note that the notation used is that of the generalised model of~\cite{Bonifacio_propagation} which, in addition to the FEL interaction, also describes the Collective Atomic Recoil Laser interaction~\cite{Bonifacio_carl}. In what follows larger values of the FEL parameter $\rho_F$ are used indicating a relatively high gain per undulator period. It is noted that while the usual Slowly Varying Envelope Approximation (SVEA)~\cite{svea1} would probably be violated, the radiation envelope approximations used in the derivation of the wave equation Eq.~(\ref{wave1}), are greatly reduced in the FEL allowing higher gains to be modeled~\cite{svea2,Brian_CSE}.

The extended QFEL model of Eqs. (\ref{schrod1}) and (\ref{wave1}) is  compared with the results of the original model of~\cite{Bonifacio_collective_lasing} by applying periodic boundary conditions to the radiation field $A$ and the electron wave $\psi$ over one radiation period equal to $4 \pi \rho_F $ in the $\bar{z}_1$ frame. The initial electron wavefunction is assumed uniform over $\bar{z}_1$ so that $\psi(\bar{z}_1,\bar{z}=0)=1$ , and a initial weak radiation `seed' field of $A(\bar{z}_1,\bar{z}=0) = 0.01$. A fourier decomposition can also be  applied to the electron wavefunction to recover the electron momentum states~\cite{Bonifacio_collective_lasing}:
\begin{equation}
\label{cn}
c_n(\bar{z})  = \frac{1}{4 \pi \rho_F}  \int_{0}^{4 \pi \rho_F} \psi(\bar{z}_1,\bar{z}) e^{i n \frac{ \bar{z}_1}{2 \rho_F}}  d \bar{z}_1,	 
\end{equation}
where  $|c_n|^2$ represents the probability that an electron has a momentum differing from its resonant value by $p=n \hbar k$.  

\begin{figure}
{\includegraphics[width=1\textwidth]{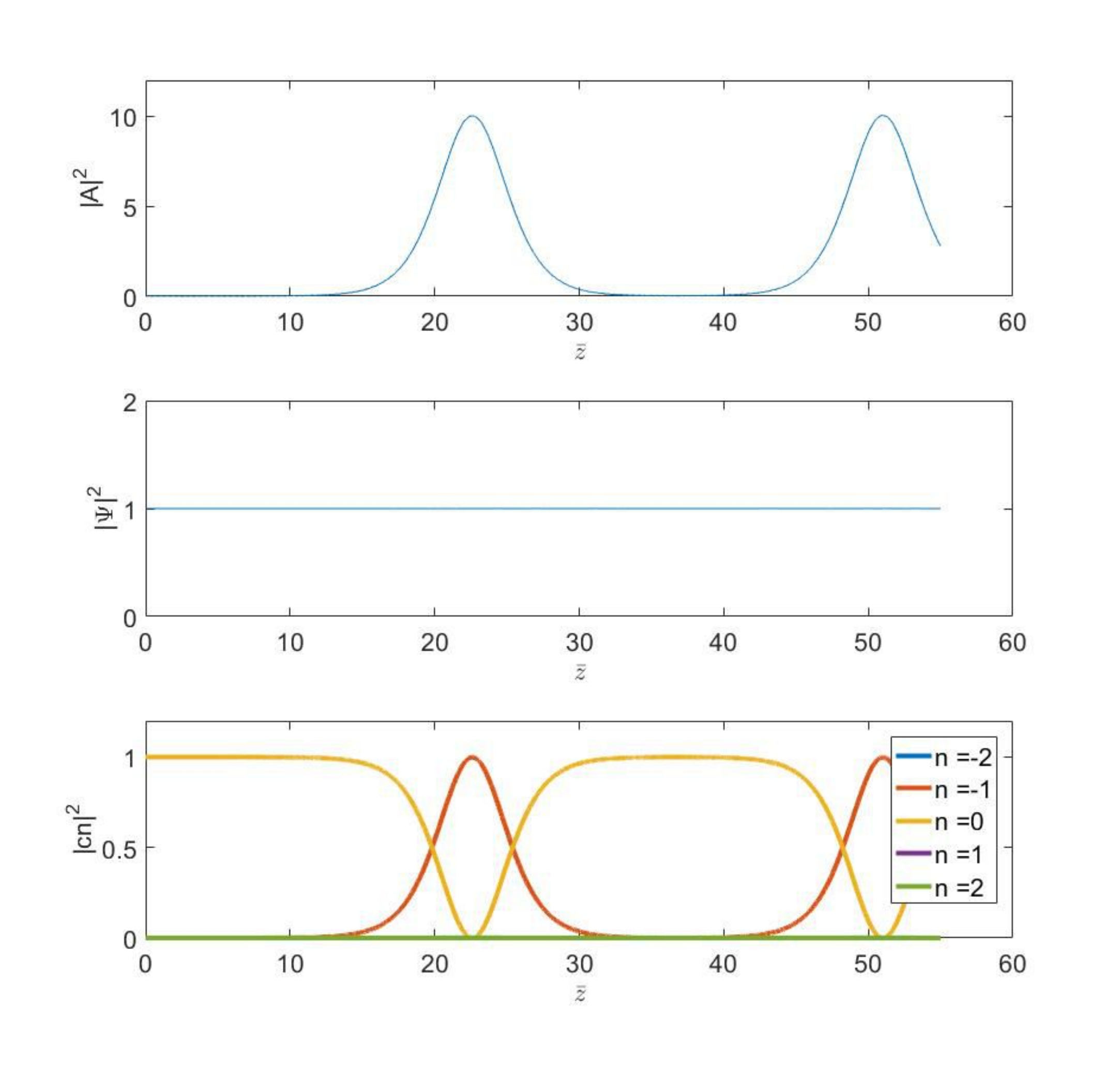}}
\caption{Steady state evolution as function of scaled propagation distance $\bar{z}$  through the undulator of: the scaled radiation intensity $|A|^2$ (top); the electron wave function density $|\psi|^2$ (middle), seen to remain constant due the conservation of electron number; and the electron momentum state occupation probability $|c_n|^2$ (bottom), for  $\bar{\rho}=0.2$, $\delta=1$ and $\rho_F=0.05$. The variables plotted are the mean values over an interval in $\bar{z}_1$ of $4\pi\rho_F$,  corresponding to one resonant radiation wavelength over which periodic boundary conditions were applied.}
\label{fig1}
\end{figure}

Figure~\ref{fig1} shows the results of a numerical solution to  Eqs. (\ref{schrod1}) and (\ref{wave1}) when these periodic boundary conditions are applied and with similar parameters to~\cite{Bonifacio_collective_lasing} (a small but different initial field of $A_0=10^{-2}$ was used here.)
It  is seen that electrons initially occupy the $n=0$ state. Transition between the $n=0$ and the $n=-1$ states, as the interaction in $\bar{z}$ develops, is seen to have excellent agreement with the original QFEL model of~\cite{Bonifacio_collective_lasing} where the electron wavefunction emits and re-absorbs a photon as the interaction progresses.  The electron wave oscillates between these two states as the radiation field grows and decays as they co-propagate through the undulator. These dynamics have also been observed in  previous QFEL models e.g.~\cite{Bonifacio_collective_lasing,Bonifacio_propagation}. This two-level dynamics can also be recovered when applying standard quantum mechanical treatments~\cite{P_Kling,P_Preiss,P_Kling2} to the Hamiltonian representation of  co-propagating radiation and electrons, with a counter-propagating wiggler field in the Bambini-Renieri frame.

\section{Pulse Propagation Effects}
\begin{figure*}
{\includegraphics[width=1\textwidth]{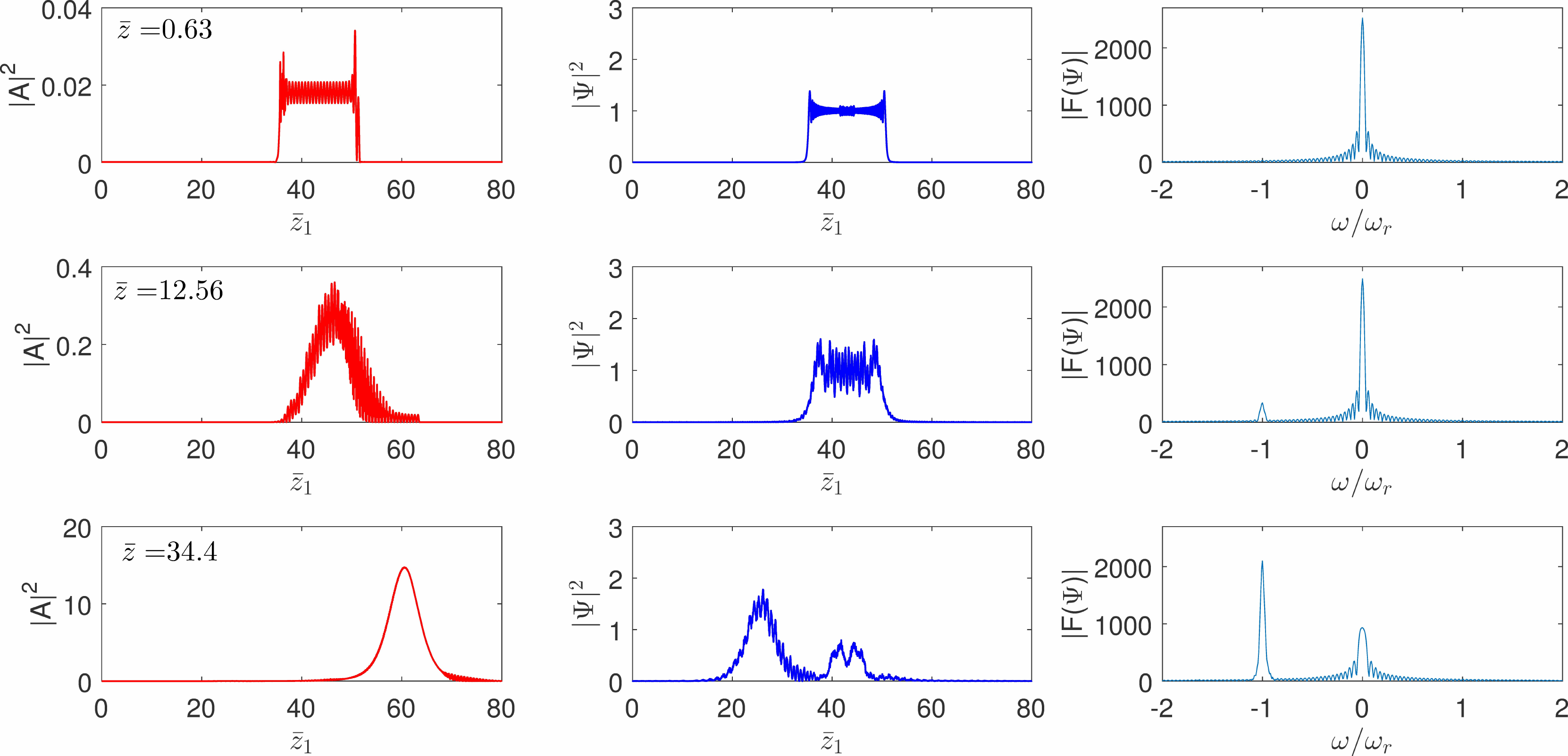}}
\caption{The scaled radiation intensity $|A|^2$, electron probability density $|\psi|^2$ and the fourier transform $|F(\psi)|$ as a function of $\bar{z}_1$ and scaled frequency $\omega/\omega_r$ for  scaled propagation distances along the undulator of Top: $\bar{z}=0.63$ (1 undulator period); Middle: $\bar{z}=12.56$ (20 undulator periods); Bottom: $\bar{z}=34.4$ (55 undulator periods). Parameters used were:  $\overline{\rho}=0.2$, $\delta=1$, $ \rho_F=0.05$ and $\overline{l_e}=16$ using an initial uniform seed field of $A_0=0.01$.}
\label{fig2}
\end{figure*}

The effects upon the  QFEL collective instability of the relative propagation of the radiation field through an electron pulse of finite dimension is now considered using the extended, coupled Maxwell -- Sch\"odinger equations of Eqs.~(\ref{schrod1}) and (\ref{wave1}). The scaled electron pulse length $l_e$, is given in units of the cooperation length $l_c$, i.e.~in units of $\bar{z}_1$, as $\bar{l}_e = l_e/l_c$. Here,  the case of a long electron pulse of many cooperation lengths is investigated in which $\bar{l}_e \gg 1$. 

An electron pulse of scaled length $\bar{l}_e=16$, FEL parameter $\rho_F=0.05$ and QFEL parameter in the quantum limit of $\bar{\rho}=0.2$ and with optimum detuning of $\delta=1$, and a small, uniform  initial radiation seed field of $A_0=0.01$ is now considered. An electron pulse with a `top-hat' current profile is used. This has a discontinuous current at either end of the pulse which, in previous classical models, has been shown to act as a strong source of Coherent Spontaneous Emission (CSE)~\cite{Brian_CSE}. CSE effects have not previously been investigated in the QFEL regime, requiring the extended model used here.
While top-hat or other current profiles with sufficiently large current gradients to generate significant CSE may not yet be achievable, progress in this field is being made~\cite{Zhengming} and cannot be ruled out in future developments. We note that the top-hat current profile used here means that, due to the current discontinuities at either ends of the pulse, there are large electron momentum uncertainties and so a large energy spread in these regions due to the Heisenberg uncertainty principle~\cite{anisimov}. Such energy spreads do not inhibit the generation of CSE however, as such emission is spontaneous and independent of the FEL interaction~\cite{Brian_CSE}.

Another effect not previously modeled is how electron recoil may change the electron pulse structure, and so how it emits radiation, by causing those electrons which have emitted a photon to drift towards the rear of the electron pulse.  

The top plots of Fig.~\ref{fig2} show (left to right) the scaled radiation power $|A|^2$, electron probability density $|\psi|^2$ and the fourier transform modulus $|F(\psi)|$ as a function of $\bar{z}_1$ and scaled frequency $\omega/\omega_r$, after approximately one undulator period ($\bar{z}\approx0.63$). The fourier transform $|F(\psi)|$ gives the momentum representation of the wavefunction and the population of the electron momentum states. On emitting a photon, an electron would then shift from $\omega/\omega_r=0$, to $\omega/\omega_r=-1$. The corresponding plots below show the same variables further into the QFEL interaction at: $\bar{z}\approx 12.56$ (20 undulator periods) and $34.4$ (55 undulator periods) respectively. 

Two peaks of radiation power are seen at either end of the electron pulse in Fig.~\ref{fig2} (top). This is the CSE due to the discontinuous change in current at $\bar{z}=0$. We note that in the classical limit of $\bar{\rho}\gg 1$, the classical theory of~\cite{Brian_CSE} predicts the peak of these radiation spikes to be $|A|^2=16\rho_F^2= 0.04$.  In the quantum limit of Fig.~\ref{fig2}, it is seen that these peak values of $|A|^2$ are slightly reduced from this value. 
This reduction is due to the quantum mechanical diffusive, second order differential term $\propto \rho_F^2/\bar{\rho}$ of Eq.~(\ref{schrod1}) which acts to diffuse and reduce any current gradients in the electron pulse. This quantum diffusive term is initially very large at the electron pulse edges, and acts to quickly reduce the large current gradient and so CSE. 
This effect of quantum diffusion of the electrons at the pulse edges after only one undulator period is also visible in the electron probability density $|\psi|^2$. This purely quantum effect cannot be modeled in a classical model or averaged QFEL models using localized periodic boundary conditions.

After sufficiently large propagation distances the effects of CSE become less significant and the underlying high gain FEL interaction becomes the dominant gain process, as seen in Fig.~\ref{fig2} (middle) for $\bar{z}\approx 12.56$. A modulation of the electron probability density $|\psi|^2$ is due to the electron bunching. Notice that the peak of the radiation emission occurs close to the centre of the electron probability density. It is noted that, in this quantum regime, the larger diffusive process reduces the electron bunching that may be achieved~\cite{Bonifacio_noise}. A further broadening of the electron pulse around its edges is also observed. The photon emission by the electrons is also seen to start populating $|F(\psi)|$ about $\omega/\omega_r=-1$. 

The macroscopic effects of electron energy loss within the pulse, following saturation of the $n=0$ to $n=-1$ transition, are clearly observed in the plots of Fig.~\ref{fig2} for the greater interaction length of $\bar{z}= 34.4$ (bottom). 
A significant fraction of the electron probability distribution has lost momentum due to the recoil of radiation emission and has drifted backwards in phase space to smaller values of $\bar{z}_1$ with a peak at $\bar{z}_1\approx 25$. The electron density and bunching are consequently reduced for $\bar{z}_1\gtrsim 40$. 
The radiation that has been emitted is propagating in near vacuum away from the electrons with increasing $\bar{z}_1$. A single pulse is observed with a peak at $\bar{z}_1\approx 61$. The electron recoil is therefore seen to further separate the electron pulse source from the radiation it emits at a rate greater than that which occurs in the previous quantum models with periodic boundary conditions, where there is no macroscopic electron transport to smaller values of $\bar{z}_1$ and only the radiation field propagates forward in $\bar{z}_1$. This increased separation rate of the recoiled electrons from the FEL interaction, a form of `self-cleaning', may make it possible to further improve the radiation pulse quality generated in the exponential gain regime. 

\begin{figure*}
{\includegraphics[width=1\textwidth]{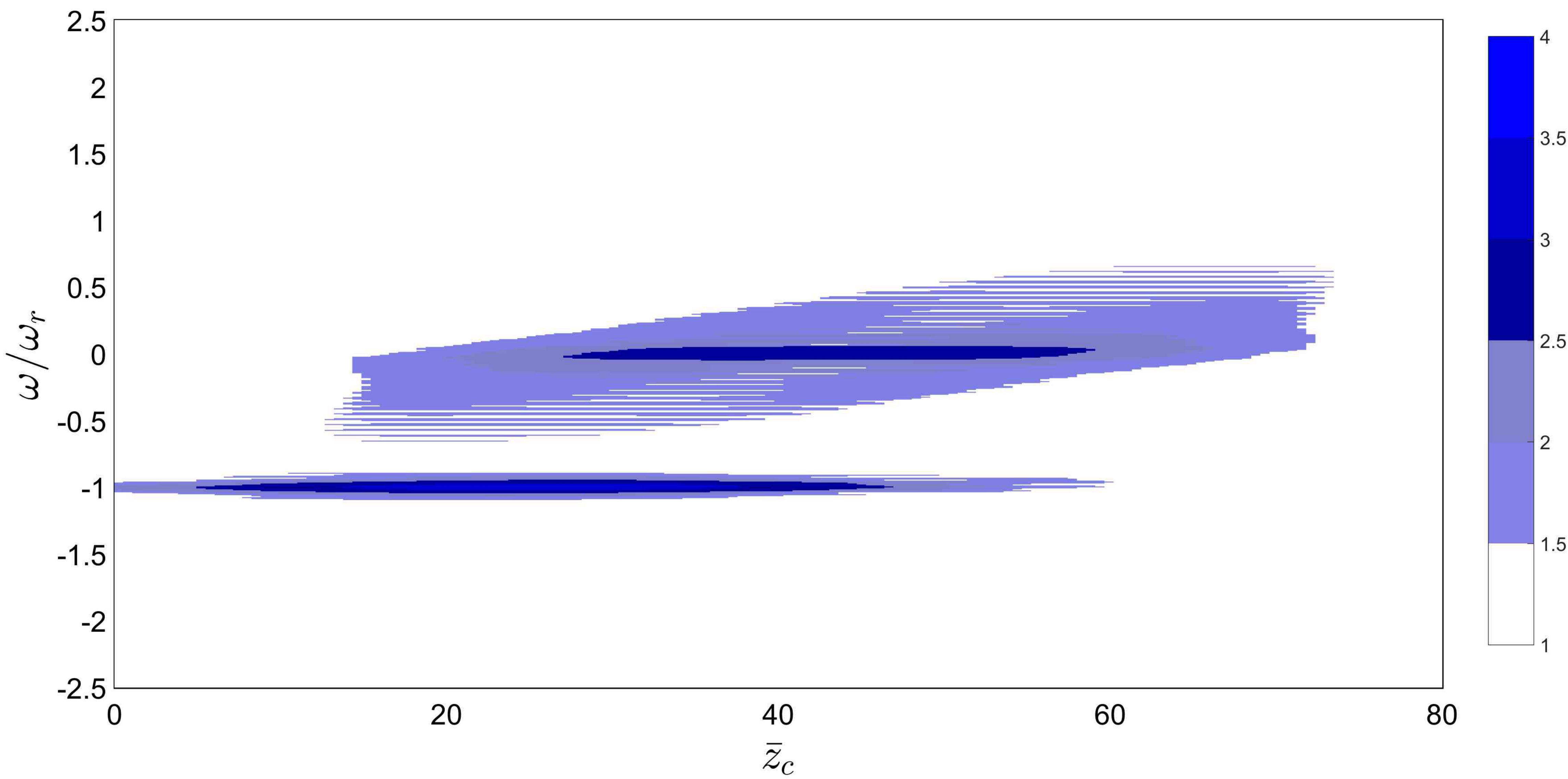}}
\caption{A Husimi representation $\log_{10}\left ( \left | H\left(\bar{z}_c, \omega \right ) \right | \right )$  of the electron wave function $\psi$ at $\bar{z}= 34.4$ for the case of Fig.~\ref{fig2} (bottom). The `self cleaning' process implicit within the extended QFEL model can be observed. That part of the electron wave function which has emitted one photon and has a scaled frequency $\omega /\omega_r = -1$, has drifted behind that part which has not emitted a photon with $\omega /\omega_r =0$. (Note that the range of $\log_{10}\left ( \left | H\left(\bar{z}_c, \omega \right ) \right | \right )$ displayed has been truncated to help contrast the regions of interest.)}
\label{fig3}
\end{figure*}

The electron dynamics of this self-cleaning regime may be better understood from a Husimi spectrogram of the electron wave function $\psi$ defined as:
\begin{equation}
H(\bar{z}_c, \omega)  = \int^{+\infty}_{-\infty} \psi(\bar{z}_1) \; e^{- \omega \bar{z}_1 } \; e^{- \frac{(\bar{z}_1 - \bar{z}_c )^2  }{ 2 \sigma_{\bar{z}_1  }^2  }  } d \bar{z}_1 
\end{equation} 
where $\sigma_{\bar{z}_1}$ defines the width of the gaussian function centred at $\bar{z}_c$ over which $\psi$ is sampled.
Figure~\ref{fig3} plots the magnitude of a Husimi transform $|H(\bar{z}_c, \omega)|$ with $\sigma_{\bar{z}_1}=10$, for the case where $\bar{z}= 34.4$ of Fig.~\ref{fig2} (bottom). It can be seen that it is those electrons that have emitted one photon that have dispersed to a position centred about $\bar{z}_1 \approx 25$ corresponding to the peak of $|\psi(\bar{z}_1)|^2$ in Fig.~\ref{fig2} (bottom). 

A comparison is now made between the extended QFEL model as presented here and the  previous model of~\cite{Bonifacio_propagation}, which is valid in the limit $\rho_F\ll 1$ and uses the notation $\bar{t}$ in place of the $\bar{z}$ used here.  In the previous model, a multiple scaling of the coordinate system is used. For each position in the pulse $\bar{z}_1$, periodic boundary conditions are applied over a ponderomotive period, corresponding to a range in $\bar{z}_1$ of $4\pi\rho_F$, and the wave equation is averaged over this interval. 
In this model of~\cite{Bonifacio_propagation}, dispersive effects acting on the electron wave function $\psi$ are confined to evolve locally within each of the ponderomotive periods modeled and cannot propagate throughout, or therefore affect, the whole pulse. Consequently, effects such as those described by the  solutions of the extended model of Figs.~\ref{fig2} and~\ref{fig3}, cannot be modeled  and the general form of the radiation pulses emitted can be expected to differ between the two models. 
\begin{figure*}
{\includegraphics[width=1\textwidth]{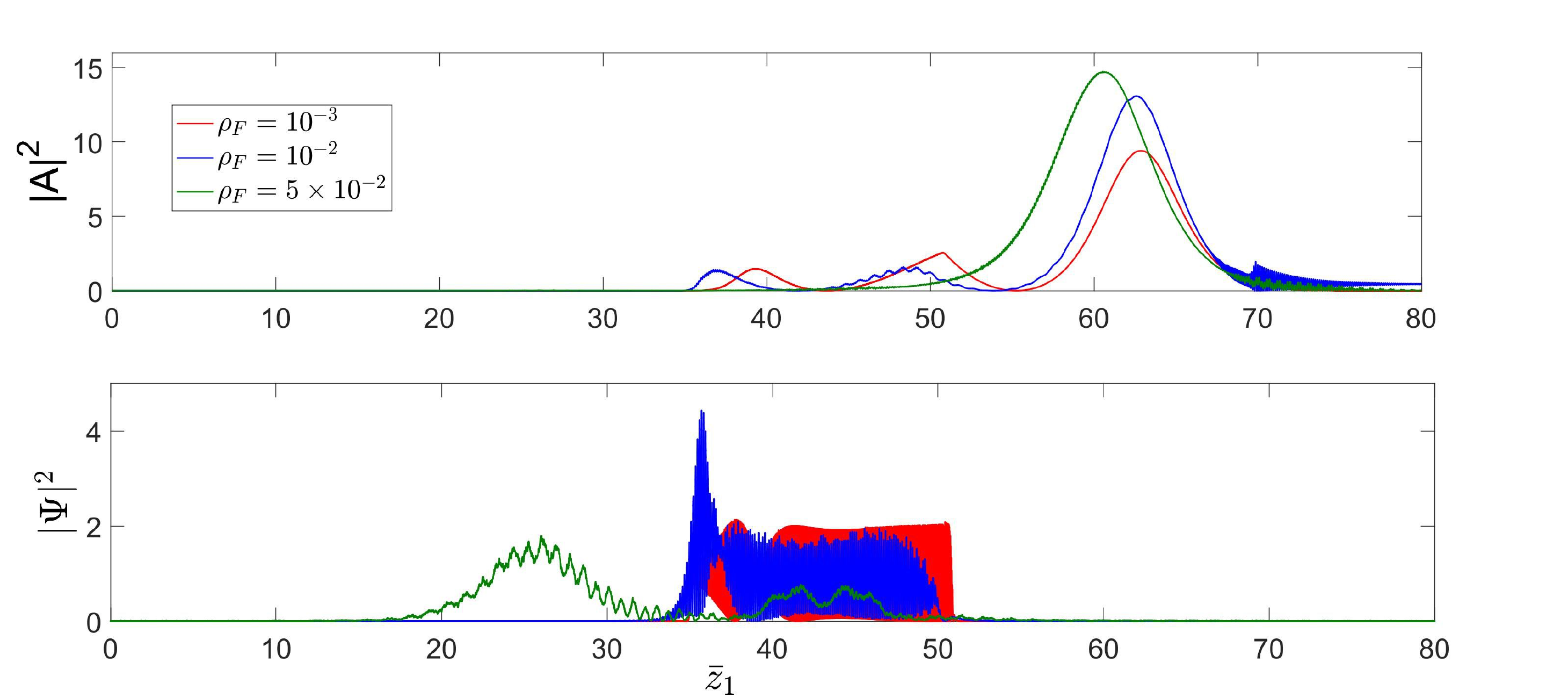}}
\caption{Top: the scaled radiation power $|A|^2$ plotted following saturation at $\bar{z}=34.4$ as a function of scaled pulse position $\bar{z}_1$ for $\rho_F=10^{-3}, 10^{-2}$ and $0.05$. 
Bottom: plots of the corresponding electron probability density.
For  small $\rho_F=10^{-3}$, there is small electron dispersion and good agreement with previous  models.
For larger $\rho_F=10^{-2}$ and $0.05$, dispersive (recoil) effects increase, enhancing the drift of electrons to smaller values of $\bar{z}_1$ and away from the emitted radiation. This drift reduces radiation re-absorption and is seen to generate a `cleaner', slightly higher power radiation pulse.}
\label{fig4}
\end{figure*}
The differences on increasing $\rho_F$ are shown in Fig.~\ref{fig4} which used the extended model of Eqs.~(\ref{schrod1}) and (\ref{wave1}) to plot the scaled radiation power $|A|^2$ and electron probability density $|\psi|^2$ in the saturation regime at $\bar{z}\approx 34.4$, as a function of position in the pulse $\bar{z}_1$. (Other parameters have the same values.) For the smaller value of  $\rho_F = 10^{-3}$, the  wavefunction of the electron pulse has little dispersion due to the $\rho_{F}^2/\overline{\rho}$ scaling of the dispersive term of~(\ref{schrod1}). It is seen that the main scaled peak power $|A|^2\lesssim 10$ and that the electron pulse emits two smaller, trailing secondary pulses.  This is in good agreement with the previous model solutions using very similar parameters as shown in Fig.~10 of~\cite{Bonifacio_propagation}. 

Significant differences are observed on increasing $\rho_F$ to $10^{-2}$ and $0.05$ of Fig.~\ref{fig4}. The electron probability density is seen to have shifted to smaller values of $\bar{z}_1$ due to the electron recoil on emission of radiation as shown in Figs.~\ref{fig2} and~\ref{fig3}. This drift of the energy depleted part of the electron density away from the radiation pulse inhibits re-absorption in a `self cleaning' process, where the recoiled electrons are removed by dispersion from the FEL interaction around the leading spike, preventing them further inhibiting the FEL interaction. This behaviour is not present in the previous model of~\cite{Bonifacio_propagation}. 

\section{Conclusion}
An extended model of a FEL operating in the quantum regime was presented which neither applies localized periodic boundary conditions nor averages the FEL interaction over these intervals. This extended QFEL model is the equivalent quantum description of the classical model of~\cite{Brian_CSE}. Both a Schr\"odinger wavefunction and momentum-state representation of the electrons was used. The extended model reduces to the previous QFEL models in the limit of the FEL parameter $\rho_F\rightarrow 0$. Both  steady-state operation and the effects of pulse propagation were investigated. The latter used quite large, as yet probably experimentally unrealizable, values of $\rho_F$ to demonstrate the principle differences between models. Application of periodic boundary conditions to the electron wavefunction and radiation field of the extended model recovered the dynamics of the previous QFEL models. 

When including pulse propagation effects in the extended QFEL model it was shown that, as in classical FEL theory, the quantum description results in CSE from the electron pulse. Like the classical FEL regime, this quantum model of CSE depends upon the electron pulse shape and is consistent with that previously observed in the classical FEL models. When operating in the quantum regime, the CSE emission can be rapidly inhibited by a quantum diffusion of the electron wavefunction in regions of high current gradient. 

For relatively short electron pulses, the recoil of electrons which have emitted radiation was shown to cause the electron pulse structure to change by causing the lower momentum states to diffuse towards the rear of the pulse. Such effects are not modeled in the previous QFEL models. The effect of this spatial separation of electrons in different momentum states is to reduce re-absorbtion of the radiation in a `self-cleaning' process which results in a better defined radiation pulse structure. Of course, these cleansing effects would be reduced in longer electron pulses. 

It is important to note that the effects described by the unaveraged QFEL model here become important only for quite large values of the FEL parameter $\rho_F$, or for shorter electron pulses.  While these values appear not to be realistically achievable using current experimental parameters~\cite{oc17}, that may change in future developments in a push towards methods of generating coherent radiation at ever shorter wavelengths. 

\section*{Funding}
We gratefully acknowledge support of the Science and Technology Facilities Council Agreement Number 4163192
Release \#3; ARCHIE-WeSt HPC, EPSRC grant EP/K000586/1; EPSRC Grant EP/M011607/1; and John von Neumann Institute for Computing (NIC) on JUROPA at J\"ulich Supercomputing Centre (JSC), under project HHH20.

\section*{Acknowledgments}
The authors thank Daniel Seipt and Gordon Robb for helpful discussions.
One of us (BWJMcN) would like to acknowledge a great personal debt to his recently departed friend and colleague Rodolfo Bonifacio, without whom this and much previous research would not have been possible.  

\section*{Disclosures}
The authors declare that there are no conflicts of interest related to this article.

\end{document}